\documentclass
	[
      a4paper,twocolumn,twoside,10pt,showkeys, 
    ]
	{revtex4}

\usepackage{amsmath,amssymb}	
\usepackage{bm}					
\usepackage{color}					
\usepackage{graphicx}				
\usepackage{dcolumn}				
\usepackage[
  colorlinks=true, citebordercolor={0 0 0}, linkcolor=blue, citecolor=blue, urlcolor=blue
]{hyperref}
\usepackage{setspace}

\newcommand{\figCapSkip}{\vspace{-4ex}}	
\makeatletter
\def\frontmatter@abstractwidth{0.9\textwidth}	
\makeatother

\usepackage{siunitx}

\begin{document}


\newcommand{\By}{$\times$}
\newcommand{\SqrtBy}[2]{$\sqrt{#1}$\kern0.2ex$\times$\kern-0.2ex$\sqrt{#2}$}
\newcommand{\Degree}{$^\circ$}
\newcommand{\DegreeC}{$^\circ$C\,}
\newcommand{\Ohmcm}{$\Omega\cdot$cm}


\title{
Nanoscale Fabrication of Graphene by Hydrogen-Plasma Etching 
}


\author{Takamoto Yokosawa}
\email[Corresponding author: ]{Takamoto.Yokosawa@anritsu.com}
\author{Masahiro Kamada}
\author{Taisuke Ochi}
\author{Yuki~Koga$^1$}
\author{Rin Takehara$^1$}
\author{Masahiro Hara$^{2, 3}$}
\author{Tomohiro Matsui}
\email[Corresponding author: ]{Tomohiro.Matsui@anritsu.com}
\affiliation{%
Advanced Research Laboratory, Anritsu Corporation, Onna 5-1-1, Atsugi, Kanagawa 243-8555, Japan
}

\affiliation{%
$^1$Graduate School of Science and Technology, Kumamoto University, Kurokami 2-39-1, Chuo-ku, Kumamoto 860-8555, Japan
}

\affiliation{%
$^2$Faculty of Advanced Science and Technology, Kumamoto University, Kurokami 2-39-1, Chuo-ku, Kumamoto 860-8555, Japan
}
\affiliation{%
$^3$Institute of Industrial Nanomaterials, Kumamoto University, Kurokami 2-39-1, Chuo-ku, Kumamoto 860-8555, Japan
}

\begin{abstract}
\vspace*{1mm}

Graphene is attracting vast interest due to its superior electronic and mechanical properties.
However, structure and electronic properties of its edge are often neglected, although they are important for nanoscale devices because the edge ratio becomes larger by decreasing the device size.
In this study, we suggest a way to fabricate a graphene with atomically aligned zigzag edges by applying hydrogen-plasma etching (HPE) technique.
By patterning a graphene prior to HPE, it is succeeded to shape a graphene in desired structure.
Both atomic force microscopy and Raman spectroscopy confirm that the graphene shaped by this technique preserves its honeycomb structure even on the edge, which is aligned with zigzag structure.
Although the mechanism of the anisotropic etching by hydrogen-plasma have not been clarified yet, the sample position dependence of the etching rate suggests that the hydrogen-radicals are responsible for the anisotropic etching.
\end{abstract}


\maketitle
\newpage


\section{Introduction}

\vspace{-0.25\baselineskip}
Graphene is a hopeful material for future electronic devices\cite{Novoselov2012}.
The electronic and mechanical properties of graphene itself, namely the bulk properties, had been studied extensively since its discovery\cite{Novoselov2004, Novoselov2005, Zhang2005}.
However, the experimental study of its edge state is rather limited, even though the understanding and controlling of the edge property is important for nanometer scale devices since the edge portion becomes larger for smaller devices.
It is especially the case for graphene because there are two types of edge structure, zigzag and armchair, with totally different electronic properties.
The electronic property of the armchair edge is roughly the same as that of bulk graphene since the bipartite symmetry is preserved.
On the other hand, the sublattice symmetry is broken along the zigzag edge and, therefore, the zigzag edge can possess a characteristic electronic state named a zigzag edge state\cite{Fujita1996}.
A flat band appears at the Fermi energy ($E_{\mathrm{F}}$), and results in an electronic state strongly localized on the edge.
Such zigzag edge states had first been confirmed around edges on graphite surfaces by scanning tunneling microscopy and spectroscopy (STM/S)\cite{NiimiMatsuiKambaraEtAl2005, NiimiMatsuiKambaraEtAl2006, KobayashiFukuiEnokiEtAl2005}, in which a single peak appears at around $E_{\mathrm{F}}$ in the electronic local density of state (LDOS) only around the edge.
Interestingly, such a zigzag edge state is expected to be spin polarized\cite{Fujita1996}.
Spins can be polarized ferromagnetically along an edge, while antiferromagnetically between edges for a zigzag graphene nanoribbon (z-GNR).
However, the experimental study for such a state is limited due to the difficulty to fabricate a zigzag edge in ideal shape.
The zigzag edge should not only be atomically precise but also be terminated by only one hydrogen (H) atom to preserve the sp$^2$ bonding.

Graphene is often micro-fabricated by either oxygen-plasma etching (OPE)\cite{Han2007, Stampfer2008} or cutting by scanning probes\cite{Puddy2011, Magda2014}.
But one cannot expect atomically aligned edges by such a top-down technique.
Moreover, it is not clear how the edges are terminated.
On the other hand, one can obtain GNRs with atomically precise and sp$^2$ bonded edges by polymerization of benzene-based molecules\cite{Talirz2013, Ruffieux2016}.
However, the size of a GNR is limited by such a bottom-up technique.

Rather recently, it is found that the ideal zigzag edge can be fabricated by etching graphene and graphite surface using H-plasma\cite{Yang2010, Diankov2013, Hug2017, Matsui2019}, in which monatomic deep hexagonal nanopits surrounded by zigzag edges are created.
The LDOS on the edge shows a sharp peak at $E_{\mathrm{F}}$ and the suppression of the LDOS next to the peak suggests that the edge shows zigzag structure with atomic precision\cite{Amend2018}.
High-resolution electron energy loss spectroscopy (HREELS) suggests that the edge is sp$^2$ bonded and terminated by only one H\cite{Ochi2022}.
It also shows that the termination is robust in ambient condition.
Moreover, STS studies show that the single peak in LDOS around an isolated zigzag edge changes to a double peak on edges of a z\nobreakdash-GNR, which strongly suggest the appearance of the spin polarized state\cite{Matsui20XX}.
Here, although the detailed mechanism of the anisotropic etching have not been clarified yet, the etching process can be understood in two steps, i.e., the H-plasma first creates defects and then enlarges these defects into nanopits.
Therefore, if the position of the defect as a nucleation center of nanopit can be controlled, one can expect to shape a graphene in desired structure with zigzag edges.
Although such technique has been suggested from a research group\cite{Yang2010, Shi2011}, it is not well established.

In this paper, we set up a H-plasma etching (HPE) system and established a way to shape a graphene by HPE.
The study of the etching behavior of our setup suggests that not only the transverse distance from the plasma glow but also a radial position inside the cylindrical chamber makes the etching results different.
The sudden change of the anisotropic etching at temperature between \SIlist{400; 500}{\degreeCelsius} shown in Reference\cite{Matsui2019} was reproduced with our setup.
By preparing nucleation center for hexagonal nanopit using CHF$_3$-plasma, we succeeded to shape a graphene into a desired structure by applying HPE. 
The atomic force microscopy (AFM) and Raman spectroscopy suggest that the graphene is flat to the very edge of the device, and the edge is zigzag structure.

\section{Experimental}
The HPE system was set following the one in Reference\cite{Matsui2019}, but the diameter of the etching chamber made by quartz tube was smaller (i.d.\,\SI{30}{\mm}) to make the cooling time shorter.
The etching parameter dependence of this HPE system was studied by observing the etched HOPG (highly oriented pyrolytic graphite) surfaces with STM.
The HOPG was typically \SI{10 x 3}{\mm} \By$^{\mathrm{t}}$\SI{1}{\mm}, and it was located about \SI{10}{\mm} from the center of the etching chamber (that was equivalent to about \SI{5}{\mm} from the quartz tube wall) when it was etched at the bottom of the chamber.
Note that, a graphite sheet and a glass plate were used as a sample stage inside the HPE chamber, and no obvious differences were observed depending on the stage material.

For the nano-fabrication, on the other hand, we used graphene prepared by the exfoliation of either HOPG or Kish graphite on SiO$_2$($^{\mathrm{t}}$\SI{285}{\nm})/Si substrate ($\sim$\SI{10 x 10}{\mm}).
The etching by CHF$_3$-plasma and oxygen-plasma were performed using commercial reactive ion etching system (RIE-10NR, Samco) and plasma cleaner (PDC-32G, Harrick Plasma), respectively.
The resultant surfaces were observed by peak-force tapping mode AFM (Dimension XR, Bruker), and the surface crystalline structures were studied by Raman spectroscopy (inVia Raman microscope, Renishaw plc.).

\section{Results and Discussion}

\subsection{Etching parameter dependence of HPE}

\begin{figure}[tb]
\begin{center}
\includegraphics[
  width=\columnwidth,
]{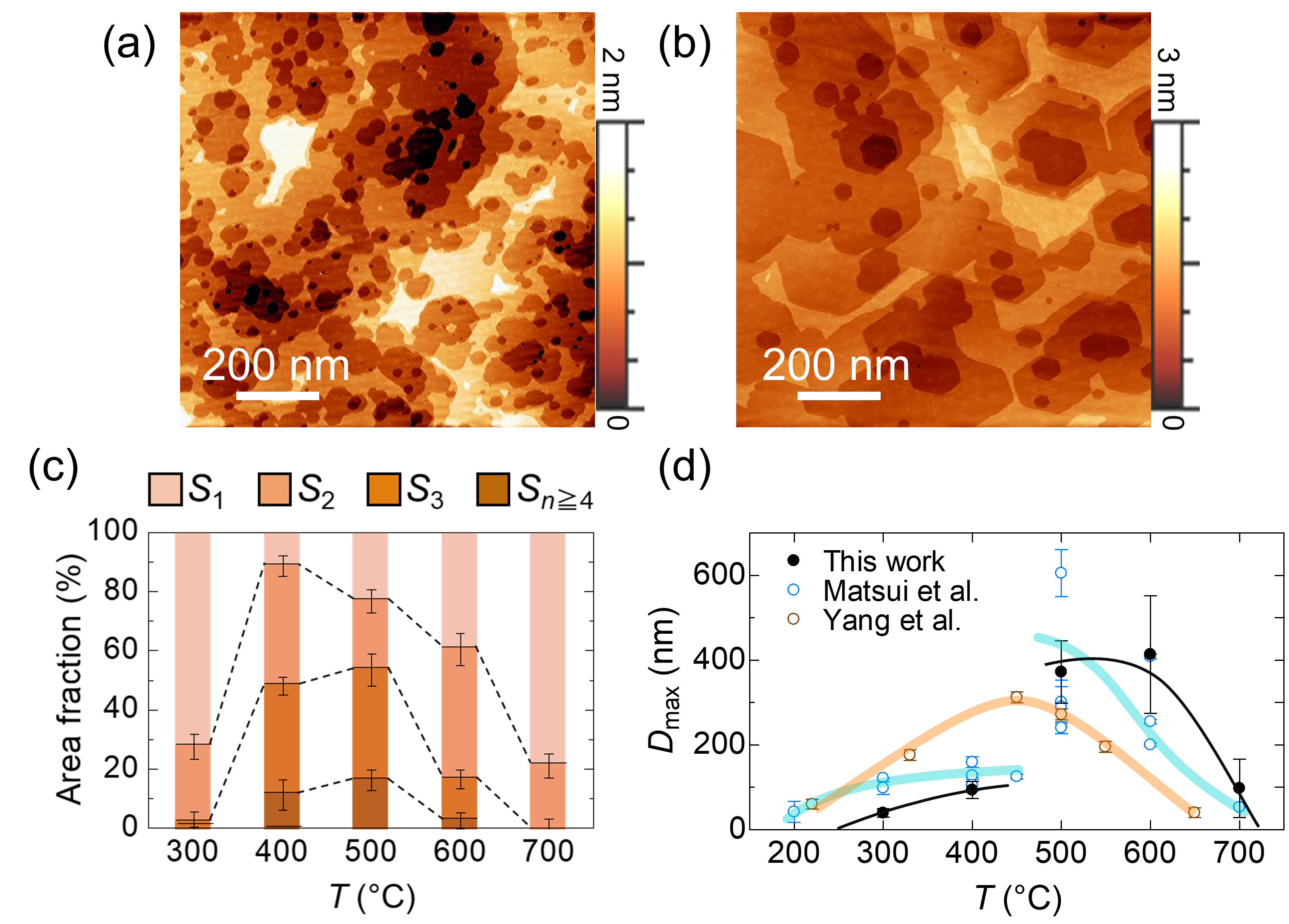}
\end{center}\figCapSkip
\caption{
  \label{fig_1}
  (a)(b) STM images of HOPG etched using H\protect\nobreakdash-plasma at $T =$ \SI{400}{\degreeCelsius} (a) and \SI{500}{\degreeCelsius} (b).
  (c)(d) $T$ dependence of $S_{\mathrm{n}}$ (c) and $D_{\mathrm{max}}$ (d).
  The data from References\cite{Matsui2019, Yang2010} are scaled to be qualitatively comparable to the data in this work.
  The solid lines in (d) are drawn for guide to the eye. 
  The etching parameters other than $T$ are kept at $P = \SI{1.5e+2}{\Pa}$, $W_{\mathrm{RF}} = \SI{20}{W}, t = \SI{5}{\minute}.$
}
\end{figure}

The shape of the nanopit created by HPE depends strongly on the etching parameters such as etching temperature ($T$), H$_2$ pressure ($P$), plasma excitation power ($W_{\mathrm{RF}}$) and the distance ($l$) from the plasma glow to the sample\cite{Matsui2019, Yang2010, Hug2017, Diankov2013}.
The etching occurs in certain $T$ range but the detailed behavior is different between research groups.
Some suggest smooth change of the etching rate\cite{Yang2010, Diankov2013}, while other suggest that the etching behavior drastically changes at intermediate $T$\cite{Matsui2019}.
The $T$ dependence of our setup is confirmed in Figure~\ref{fig_1}.
Figure~\ref{fig_1}(a,\,b) show the STM images of the typical HOPG surfaces etched at $T =$ \SIlist{400; 500}{\degreeCelsius}, respectively.
They clearly suggest that the etching behavior changes drastically at narrow $T$ window between \SIlist{400; 500}{\degreeCelsius}.
This change is quantitatively shown by calculating the surface ratio of $n$-th layer from the surface ($S_{\mathrm{n}}$) and the diameter of the nanopit ($D_{\mathrm{max}}$).
Here, the definition of $S_{\mathrm{n}}$ and $D_{\mathrm{max}}$ follow those in Reference\cite{Matsui2019}.
The $T$ dependence of $S_{\mathrm{n}}$ and $D_{\mathrm{max}}$ are shown in Figure~\ref{fig_1}(c,\,d), respectively.
They again show that the etching behavior changes between \SIlist{400; 500}{\degreeCelsius}, similarly to the case in Reference\cite{Matsui2019} but different from References\cite{Yang2010, Diankov2013}.
Circular (isotropic etching) and small nanopits are created at $T$ lower than \SI{400}{\degreeCelsius}, while hexagonal (anisotropic etching) and relatively large nanopits are created at $T$ higher than \SI{500}{\degreeCelsius}.
The reason why the etching behavior changes drastically at such intermediate $T$ is still unclear, but this observation suggests that such change is not an artifact and some mechanism is surely existing.

\begin{figure}[tbh]
\begin{center}
\includegraphics[
  width=0.94\columnwidth
]{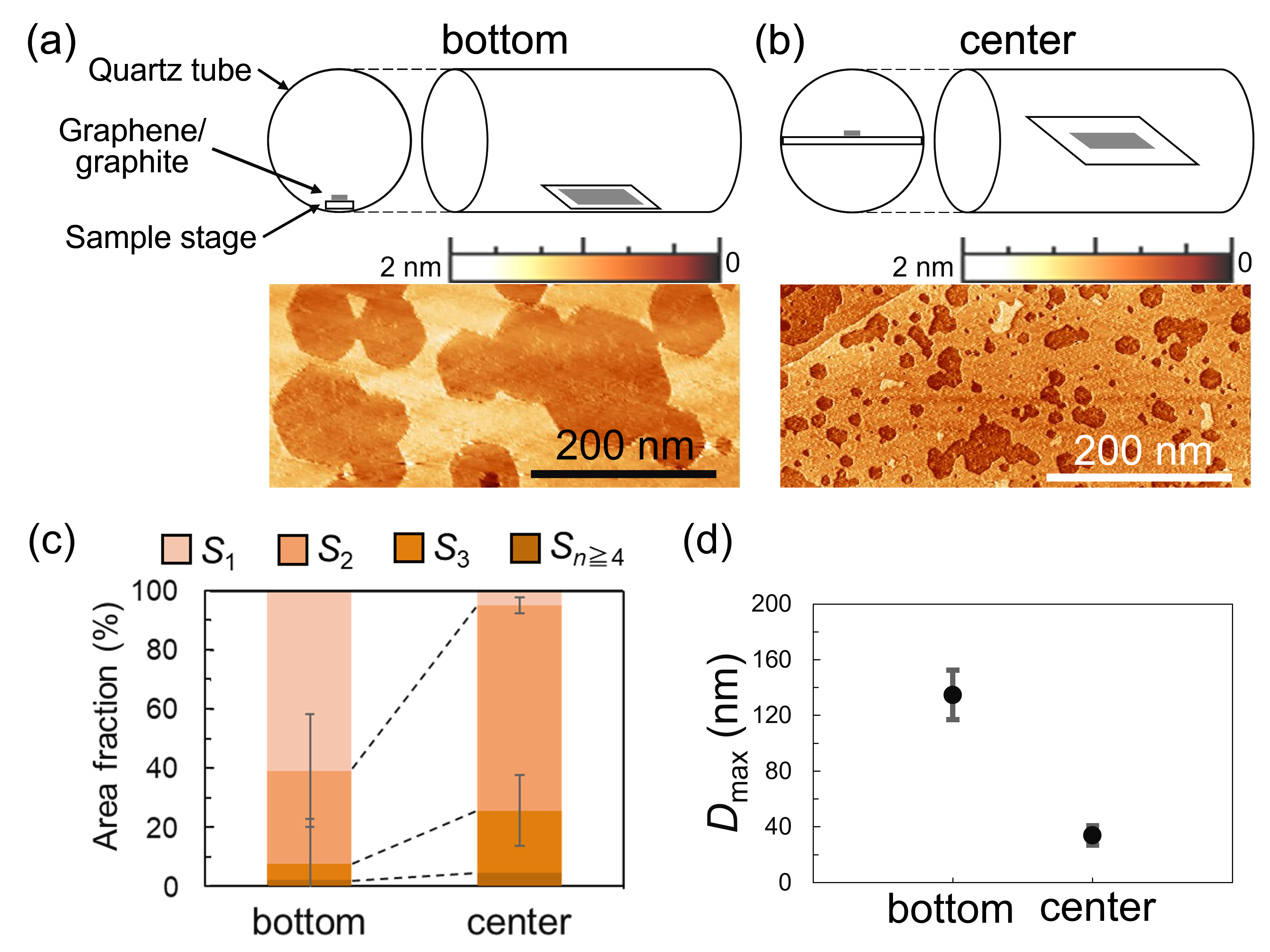}
\end{center}\figCapSkip
\caption{
  \label{fig_2}
  (a)(b) STM images of HOPG etched using H\protect\nobreakdash-plasma at the bottom (a) and at the center (b) of the chamber.
  Upper side of each image shows schematic of the position in the chamber.
  (c)(d) Position dependence of $S_{\mathrm{n}}$ (c) and $D_{\mathrm{max}}$ (d).
  The etching parameters are kept at $T = \SI{600}{\degreeCelsius}$, $P = \SI{1.5e+2}{\Pa}$, $W_{\mathrm{RF}} = \SI{20}{W}, t = \SI{5}{\minute}.$
}
\end{figure}

In addition, we found that the etching can also be different at where the sample is set radially in the cylindrical quartz tube chamber.
Figure~\ref{fig_2}(a,\,b) show the typical STM images of HOPG surfaces etched at the bottom and at the center of the chamber, respectively.
The nanopits are small and circular when HOPG is etched at the center, while they are large and hexagonal at the bottom, even though the etching parameters are the same.
The $S_{\mathrm{n}}$ and $D_{\mathrm{max}}$ for two positions are shown in Figure~\ref{fig_2}(c,\,d), respectively.
These clearly show that the etching effect is weaker but the nanopit size is about three times larger when HOPG is etched at the bottom than at the center.

Such change of the etching can be understood by considering the radial variation of H ions and H radicals.
Since H ions, such as H$^+$, H$_2^+$, H$_3^+$, can decay faster than H radicals, the density of H radicals can be higher than that of H ions near the quartz wall.
It suggests that the H ions are responsible for the defect creation, while H radicals are for the anisotropic etching of the defects.
This scenario is consistent to the other measurements showing $P$\cite{Hug2017, Matsui2019}, $W_{\mathrm{RF}}$\cite{Matsui2019} and $l$\cite{Hug2017} dependences of HPE.

\subsection{Nano-fabrication of graphene using HPE}

\begin{figure}[tb]
\begin{center}
\includegraphics[
  width=\columnwidth
]{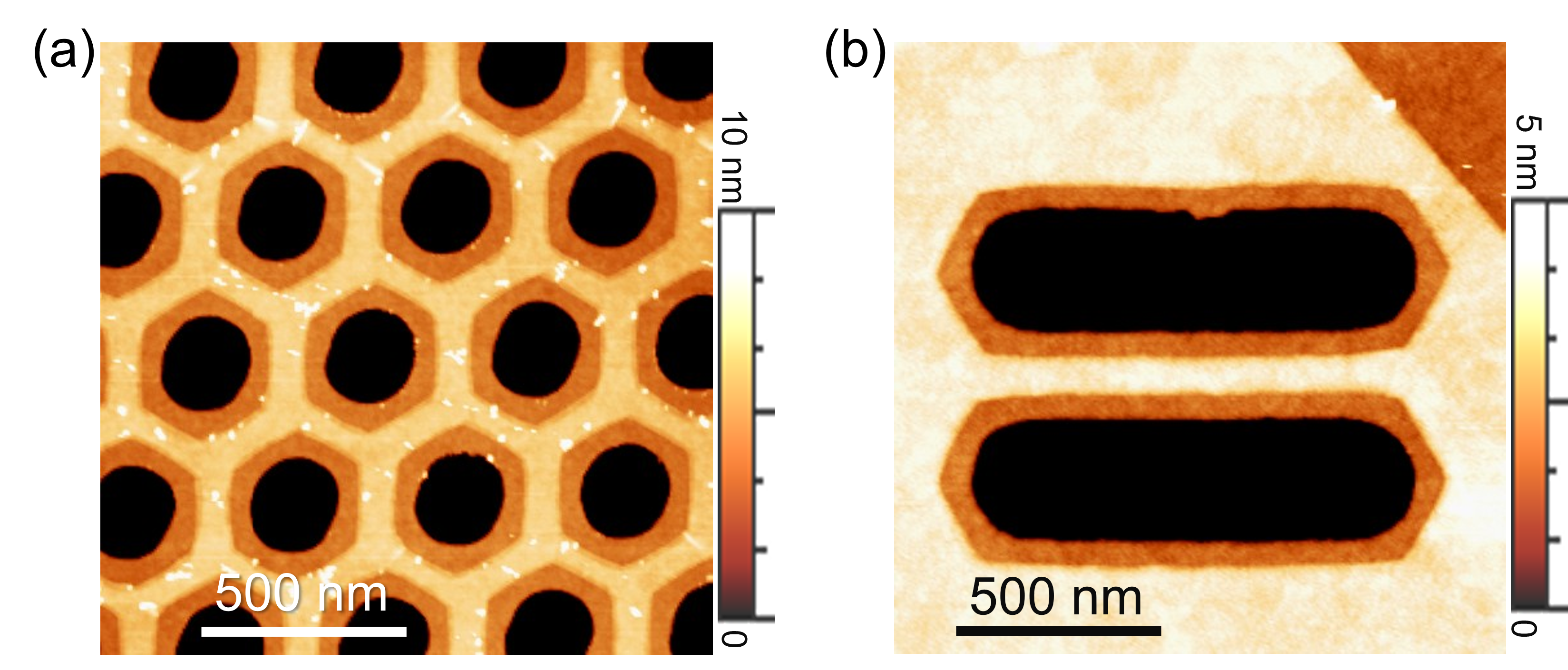}
\end{center}\figCapSkip
\caption{\label{fig_3}
  AFM images of (a) z-GNR network (\SI{2.3}{\nm} thick) and (b) single z-GNR (\SI{2.1}{\nm} thick) obtained between two elongated hexagonal nanopits.
  The brightest, second brightest and darkest region correspond to graphene, SiO$_2$ substrate and pre-patterned holes, respectively. 
  The etching parameters for HPE are $T = \SI{600}{\degreeCelsius}$, $P = \SI{1.5e+2}{\Pa}$, $W_{\mathrm{RF}} = \SI{20}{W}, t = \SI{1}{\minute}$ for both images.
}
\end{figure}

To fabricate a graphene by HPE, we first followed the way reported previously\cite{Yang2010, Shi2011}, in which a graphene is patterned by electron-beam lithography and OPE before HPE.
However, the pre-patterned circular nanopits were not enlarged hexagonally with our setups.
This is probably because the defects created by OPE cannot be nucleation centers for the anisotropic etching by HPE.
Similarly, the anisotropic etching does not start from step edges of a graphite and edges of a graphene flake.
A graphene flake just becomes small, and the flake edge is not shaped into zigzag by HPE, even though hexagonal nanopits are created on the interior of the flake.
For instance, it can be observed in Figure~\ref{fig_3}(b) and Figure~\ref{fig_5}(a) that the angle between the flake edge and the hexagonal nanopit, the edge of which is aligned into zigzag, does not a multiple of \ang{60}, therefore, the flake edge is not aligned into zigzag.
These facts suggest that, not all the defects cannot be nucleation centers for the anisotropic etching by HPE.

In this study, we succeeded to fabricate a graphene by HPE the pre-patterned nucleation center which was prepared by CHF$_3$-plasma etching (CPE) instead of OPE.
Figure~\ref{fig_3}(a) shows an AFM image of a graphene in which hexagonal nanopits are aligned triangularly.
Since the CHF$_3$-plasma can etch not only graphene but also SiO$_2$, deep holes are also created on the substrate.
The deep circular holes at the center of the hexagonal nanopits are the pre-patterned holes created by CPE.
In this case, a network of z\nobreakdash-GNR with about \SI{70}{\nm} wide and \SI{170}{\nm} long is successfully fabricated.

Moreover, the hexagonal nanopit will be elongated if the pre-patterned hole is ellipse as shown in Figure~\ref{fig_3}(b).
This fact means that even a long z-GNR can be fabricated between two elongated hexagonal nanopits.
For instance, a z-GNR with about \SI{100}{\nm} wide and \SI{1}{\um} long is obtained in Figure~\ref{fig_3}(b).
Since the etching amount can be controlled by etching time ($t$) at fixed $T, P$ and $W_{\mathrm{RF}}$, one can fabricate a graphene into desired shape and size.
The HPE is anisotropic for graphene more than two layers with our setups, similarly to References.\cite{Diankov2013, Hug2017} but different from Reference\cite{Yang2010}, which make it difficult to obtain a monolayer graphene device.
However, it is better to note that multilayer graphene has variety of electronic and mechanical properties depending on the number of layers and that even a thick graphene can be fabricated by this technique as far as the pre-patterned hole penetrates the graphene.

\subsection{Evaluation of the edges}

\begin{figure*}[tbh]
\begin{center}
\includegraphics[
  width=1.8\columnwidth
]{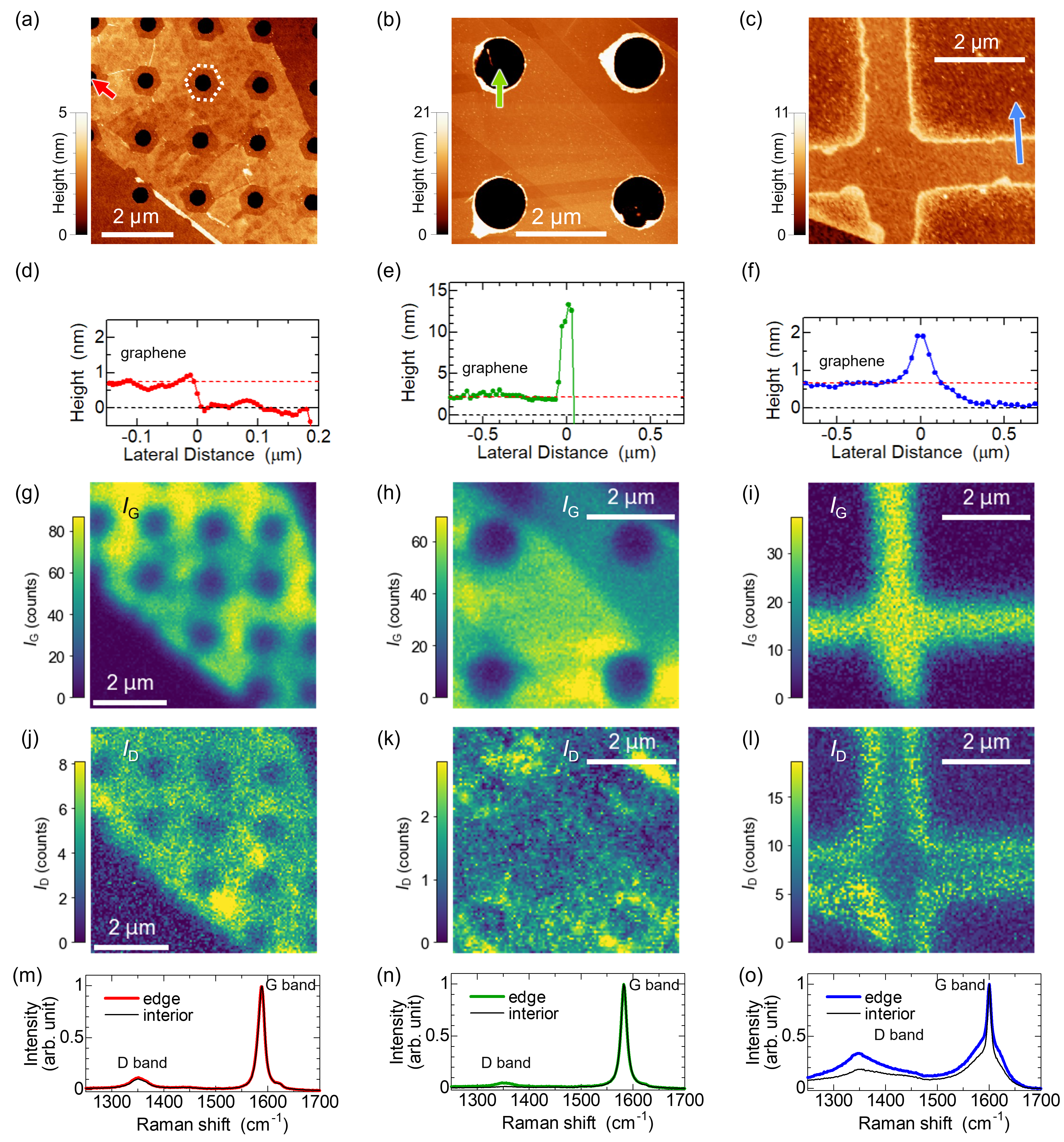}
\end{center}\figCapSkip
\caption{
  \label{fig_5}
  (a)--(c) AFM images of the graphene fabricated by HPE (a), CPE (b) and OPE (c).
  (d)--(f) Line profiles along the arrows shown in (a)--(c), respectively.
  The height is scaled from SiO$_2$ substrate surface, while the lateral distance is scaled from the edge of graphene (d) and the peak of the elevated region (e)(f).
  (g)--(l) Spatial distribution of $I_{\mathrm{G}}$ (g)--(i) and $I_{\mathrm{D}}$ (j)--(l) for the graphene fabricated by HPE (g)(j), CPE (h)(k) and OPE (i)(l), respectively.
  (m)--(o) Raman spectra averaged in the measurements shown in (g)--(l) for the graphene fabricated by HPE (m), CPE (n) and OPE (o), respectively.
  Colored (Black) line shows the spectrum averaged on edge (interior) of the shaped graphene. 
  The spectra are normalized with $I_{\mathrm{G}}$.
  The parameters for HPE are $T = \SI{600}{\degreeCelsius}$, $P = \SI{1.5e+2}{\Pa}$, $W_{\mathrm{RF}} = \SI{20}{W}, t = \SI{5}{\minute}.$
}
\end{figure*}

The quality of the edge created by HPE is compared to the ones prepared by CPE and OPE.
Figure~\ref{fig_5}(a--c) show the topographic images patterned by HPE, CPE and OPE, respectively.
Nanopits are created periodically by HPE (Figure~\ref{fig_5}(a)) and CPE (Figure~\ref{fig_5}(b)), while a graphene is shaped into crosswise by OPE (Figure~\ref{fig_5}(c)).
Although the graphene can be shaped into desired structure by any etching technique, the edges are observed differently.
The edges are elevated when graphene is etched by CPE and OPE.
The line profile across the edge (Figure~\ref{fig_5}(e,\,f)) shows that it is elevated about \SI{10}{\nm} (\SI{1.5}{\nm}) over \SI{0.1}{\um} (\SI{0.3}{\um}) wide for the graphene fabricated by CPE (OPE).
Because such a feature is not observed on SiO$_2$ substrate, one can say that graphene, rather than SiO$_2$, is elevated. 
And because the edge of the pre-patterned hole is located in the elevated region under graphene, graphene covers the edge.
Therefore, unfortunately, the exact structure and location of graphene edge is not clear in these measurements for CPE and OPE.
It should be noted, however, that the reason of these elevations is not clear. The edge of graphene can be elevated either due to the etching or to the lift-off process after the etching.
On the other hand, the graphene fabricated by HPE is flat to the edge and the boundary between graphene and SiO$_2$ is clear as can be seen in Figure~\ref{fig_5}(d).
This fact confirms that the edge created by HPE is topographically smooth.

The crystallographic structure of the edge is evaluated by Raman spectroscopy.
The spatial variations of the G (D) band intensity $I_{\mathrm{G}}$ ($I_{\mathrm{D}}$) obtained in the same area in Figure~\ref{fig_5}(a--c) are shown in Figure~\ref{fig_5}(g--i)((j--l)).
Although the $I_{\mathrm{G}}$ follows the patterned structure for every sample, $I_{\mathrm{D}}$ becomes stronger along the edge than the interior for the graphene fabricated by CPE and OPE.
For the graphene fabricated by HPE, on the other hand, $I_{\mathrm{D}}$ shows basically the same spatial distribution as $I_{\mathrm{G}}$.
As can be seen in Figure~\ref{fig_5}(c), the original shape of a graphene flake is sometimes observed in a topographic image even after the etching.
However, because no G band appears in etched area (Figure~\ref{fig_5}(i)), the graphene is etched away at least in micrometer scale.

The Raman spectra averaged in the area about \SI{400}{\nm} away from the edge (``interior'') and the area within about \SI{400}{\nm} from the edge (``edge'') for each measurement (Figure~\ref{fig_5}(g--l)) are shown in Figure~\ref{fig_5}(m--o).
Here, the edge locations are determined from the optical images simultaneously obtained with Raman spectra and the AFM images.
The D band becomes strong around the edge for the graphene fabricated by CPE and OPE, while spectra on the edge and the interior are identical for the graphene fabricated by HPE.
Considering that the D band is originated from inter-valley scattering, these results suggest that the edges prepared by CPE and OPE are defective.
In addition, since a zigzag edge does not cause inter-valley scattering, the Raman spectrum also support the fact that the edge formed by HPE is mainly zigzag structure.
Note that graphene interior are also etched by HPE since it is exposed to H-plasma, while interior is protected by resist for electron beam lithography during CPE and OPE.
Because the graphene fabricated by HPE is etched a bit long time to make the hexagonal nanopit large enough to study with Raman spectroscopy, the spatial resolution of which is about \SI{500}{\nm}, a finite amount of defects is also created and $I_{\mathrm{D}}$ appears even away from the edge as can be seen in both topographic image (Figure~\ref{fig_5}(a)) and Raman spectrum (Figure~\ref{fig_5}(m)).
However, what is important here is that the defect density and the crystalline structure is uniform inside the shaped graphene.
The $I_{\mathrm{D}}$ can be much smaller by making etching time shorter for nanoscale devices.

\section{Summary}
\vspace*{\baselineskip}
The $T$ and radial position dependences of our HPE setup were shown first in this paper.
The $T$ dependence shows a sudden jump in $D_{\mathrm{max}}$ at between \SIlist{400; 500}{\degreeCelsius} similarly to the case in Reference\cite{Matsui2019}.
It may provide some hint to understand the etching mechanism.
In addition, it is also found that the effect of the anisotropic etching (defect creation) is stronger (weaker) at the bottom of the chamber than at the center.
This fact suggests that the radial distribution of the plasma component, such as H radicals and H ions, make the difference of the etching.
Considering that the H ions are stronger radial distribution than H radicals, the anisotropic etching can be attributed to the H radicals, while the defect creation can be to the H ions.

The way of the pre-patterning is found to be crucial for the nano-fabrication of graphene by HPE.
In this study, we succeeded to shape a graphene with hexagonal nanopits using CHF$_3$-plasma as an etchant for the pre-patterning.
Importantly, the honeycomb structure of graphene is preserved even on the edge in the shaped graphene, and the edge is arranged in the zigzag structure with atomic precision.
It can be a great advantage in both fundamental researches and applications of graphene to nanoscale devices.
This reproducible way to fabricate a z-GNR can pave the way for experimental study of the spin polarized zigzag edge state.
At the same time, one can design a graphene nanoscale device more reliably with this fabrication technique because the edge of the patterned graphene is well controlled.
Moreover, a new graphene device using high electronic density of state localized around a zigzag edge can also be expected.

\begin{acknowledgments}
A part of this work was supported by ``Nanotechnology Platform Japan'' of the Ministry of Education, Culture, Sports, Science and Technology (MEXT) Grant Number JPMXP09F21UT0045, and fabrication was conducted in Takeda Cleanroom with help of Nanofabrication Platform Center of School of Engineering, the University of Tokyo, Japan.
The authors acknowledge the free-to-use software, WSxM\cite{WSxM}.
\end{acknowledgments}


\begin{thebibliography}{23}%
\makeatletter
\providecommand \@ifxundefined [1]{%
 \@ifx{#1\undefined}
}%
\providecommand \@ifnum [1]{%
 \ifnum #1\expandafter \@firstoftwo
 \else \expandafter \@secondoftwo
 \fi
}%
\providecommand \@ifx [1]{%
 \ifx #1\expandafter \@firstoftwo
 \else \expandafter \@secondoftwo
 \fi
}%
\providecommand \natexlab [1]{#1}%
\providecommand \enquote  [1]{``#1''}%
\providecommand \bibnamefont  [1]{#1}%
\providecommand \bibfnamefont [1]{#1}%
\providecommand \citenamefont [1]{#1}%
\providecommand \href@noop [0]{\@secondoftwo}%
\providecommand \href [0]{\begingroup \@sanitize@url \@href}%
\providecommand \@href[1]{\@@startlink{#1}\@@href}%
\providecommand \@@href[1]{\endgroup#1\@@endlink}%
\providecommand \@sanitize@url [0]{\catcode `\\12\catcode `\$12\catcode
  `\&12\catcode `\#12\catcode `\^12\catcode `\_12\catcode `\%12\relax}%
\providecommand \@@startlink[1]{}%
\providecommand \@@endlink[0]{}%
\providecommand \url  [0]{\begingroup\@sanitize@url \@url }%
\providecommand \@url [1]{\endgroup\@href {#1}{\urlprefix }}%
\providecommand \urlprefix  [0]{URL }%
\providecommand \Eprint [0]{\href }%
\providecommand \doibase [0]{https://doi.org/}%
\providecommand \selectlanguage [0]{\@gobble}%
\providecommand \bibinfo  [0]{\@secondoftwo}%
\providecommand \bibfield  [0]{\@secondoftwo}%
\providecommand \translation [1]{[#1]}%
\providecommand \BibitemOpen [0]{}%
\providecommand \bibitemStop [0]{}%
\providecommand \bibitemNoStop [0]{.\EOS\space}%
\providecommand \EOS [0]{\spacefactor3000\relax}%
\providecommand \BibitemShut  [1]{\csname bibitem#1\endcsname}%
\let\auto@bib@innerbib\@empty
\bibitem [{\citenamefont {Novoselov}\ \emph {et~al.}(2012)\citenamefont
  {Novoselov}, \citenamefont {Fal'ko}, \citenamefont {Colombo}, \citenamefont
  {Gellert}, \citenamefont {Schwab},\ and\ \citenamefont
  {Kim}}]{Novoselov2012}%
  \BibitemOpen
  \bibfield  {author} {\bibinfo {author} {\bibfnamefont {K.~S.}\ \bibnamefont
  {Novoselov}}, \bibinfo {author} {\bibfnamefont {V.~I.}\ \bibnamefont
  {Fal'ko}}, \bibinfo {author} {\bibfnamefont {L.}~\bibnamefont {Colombo}},
  \bibinfo {author} {\bibfnamefont {P.~R.}\ \bibnamefont {Gellert}}, \bibinfo
  {author} {\bibfnamefont {M.~G.}\ \bibnamefont {Schwab}},\ and\ \bibinfo
  {author} {\bibfnamefont {K.}~\bibnamefont {Kim}},\ }\href
  {https://doi.org/10.1038/nature11458} {\bibfield  {journal} {\bibinfo
  {journal} {Nature}\ }\textbf {\bibinfo {volume} {490}},\ \bibinfo {pages}
  {192} (\bibinfo {year} {2012})}\BibitemShut {NoStop}%
\bibitem [{\citenamefont {Novoselov}\ \emph {et~al.}(2004)\citenamefont
  {Novoselov}, \citenamefont {Geim}, \citenamefont {Morozov}, \citenamefont
  {Jiang}, \citenamefont {Zhang}, \citenamefont {Dubonos}, \citenamefont
  {Grigorieva},\ and\ \citenamefont {Firsov}}]{Novoselov2004}%
  \BibitemOpen
  \bibfield  {author} {\bibinfo {author} {\bibfnamefont {K.~S.}\ \bibnamefont
  {Novoselov}}, \bibinfo {author} {\bibfnamefont {A.~K.}\ \bibnamefont {Geim}},
  \bibinfo {author} {\bibfnamefont {S.~V.}\ \bibnamefont {Morozov}}, \bibinfo
  {author} {\bibfnamefont {D.}~\bibnamefont {Jiang}}, \bibinfo {author}
  {\bibfnamefont {Y.}~\bibnamefont {Zhang}}, \bibinfo {author} {\bibfnamefont
  {S.~V.}\ \bibnamefont {Dubonos}}, \bibinfo {author} {\bibfnamefont {I.~V.}\
  \bibnamefont {Grigorieva}},\ and\ \bibinfo {author} {\bibfnamefont {A.~A.}\
  \bibnamefont {Firsov}},\ }\href {https://doi.org/10.1126/science.1102896}
  {\bibfield  {journal} {\bibinfo  {journal} {Science}\ }\textbf {\bibinfo
  {volume} {306}},\ \bibinfo {pages} {666} (\bibinfo {year}
  {2004})}\BibitemShut {NoStop}%
\bibitem [{\citenamefont {Novoselov}\ \emph {et~al.}(2005)\citenamefont
  {Novoselov}, \citenamefont {Geim}, \citenamefont {Morozov}, \citenamefont
  {Jiang}, \citenamefont {Katsnelson}, \citenamefont {Grigorieva},
  \citenamefont {Dubonos},\ and\ \citenamefont {Firsov}}]{Novoselov2005}%
  \BibitemOpen
  \bibfield  {author} {\bibinfo {author} {\bibfnamefont {K.~S.}\ \bibnamefont
  {Novoselov}}, \bibinfo {author} {\bibfnamefont {A.~K.}\ \bibnamefont {Geim}},
  \bibinfo {author} {\bibfnamefont {S.~V.}\ \bibnamefont {Morozov}}, \bibinfo
  {author} {\bibfnamefont {D.}~\bibnamefont {Jiang}}, \bibinfo {author}
  {\bibfnamefont {M.~I.}\ \bibnamefont {Katsnelson}}, \bibinfo {author}
  {\bibfnamefont {I.~V.}\ \bibnamefont {Grigorieva}}, \bibinfo {author}
  {\bibfnamefont {S.~V.}\ \bibnamefont {Dubonos}},\ and\ \bibinfo {author}
  {\bibfnamefont {A.~A.}\ \bibnamefont {Firsov}},\ }\href
  {https://doi.org/10.1038/nature04233} {\bibfield  {journal} {\bibinfo
  {journal} {Nature}\ }\textbf {\bibinfo {volume} {438}},\ \bibinfo {pages}
  {197} (\bibinfo {year} {2005})}\BibitemShut {NoStop}%
\bibitem [{\citenamefont {Zhang}\ \emph {et~al.}(2005)\citenamefont {Zhang},
  \citenamefont {Tan}, \citenamefont {Stormer},\ and\ \citenamefont
  {Kim}}]{Zhang2005}%
  \BibitemOpen
  \bibfield  {author} {\bibinfo {author} {\bibfnamefont {Y.}~\bibnamefont
  {Zhang}}, \bibinfo {author} {\bibfnamefont {Y.-W.}\ \bibnamefont {Tan}},
  \bibinfo {author} {\bibfnamefont {H.~L.}\ \bibnamefont {Stormer}},\ and\
  \bibinfo {author} {\bibfnamefont {P.}~\bibnamefont {Kim}},\ }\href
  {https://doi.org/10.1038/nature04235} {\bibfield  {journal} {\bibinfo
  {journal} {Nature}\ }\textbf {\bibinfo {volume} {438}},\ \bibinfo {pages}
  {201} (\bibinfo {year} {2005})}\BibitemShut {NoStop}%
\bibitem [{\citenamefont {Fujita}\ \emph {et~al.}(1996)\citenamefont {Fujita},
  \citenamefont {Wakabayashi}, \citenamefont {Nakada},\ and\ \citenamefont
  {Kusakabe}}]{Fujita1996}%
  \BibitemOpen
  \bibfield  {author} {\bibinfo {author} {\bibfnamefont {M.}~\bibnamefont
  {Fujita}}, \bibinfo {author} {\bibfnamefont {K.}~\bibnamefont {Wakabayashi}},
  \bibinfo {author} {\bibfnamefont {K.}~\bibnamefont {Nakada}},\ and\ \bibinfo
  {author} {\bibfnamefont {K.}~\bibnamefont {Kusakabe}},\ }\href
  {https://doi.org/10.1143/JPSJ.65.1920} {\bibfield  {journal} {\bibinfo
  {journal} {J. Phys. Soc. Jpn.}\ }\textbf {\bibinfo {volume} {65}},\ \bibinfo
  {pages} {1920} (\bibinfo {year} {1996})}\BibitemShut {NoStop}%
\bibitem [{\citenamefont {Niimi}\ \emph {et~al.}(2005)\citenamefont {Niimi},
  \citenamefont {Matsui}, \citenamefont {Kambara}, \citenamefont {Tagami},
  \citenamefont {Tsukada},\ and\ \citenamefont
  {Fukuyama}}]{NiimiMatsuiKambaraEtAl2005}%
  \BibitemOpen
  \bibfield  {author} {\bibinfo {author} {\bibfnamefont {Y.}~\bibnamefont
  {Niimi}}, \bibinfo {author} {\bibfnamefont {T.}~\bibnamefont {Matsui}},
  \bibinfo {author} {\bibfnamefont {H.}~\bibnamefont {Kambara}}, \bibinfo
  {author} {\bibfnamefont {K.}~\bibnamefont {Tagami}}, \bibinfo {author}
  {\bibfnamefont {M.}~\bibnamefont {Tsukada}},\ and\ \bibinfo {author}
  {\bibfnamefont {H.}~\bibnamefont {Fukuyama}},\ }\href
  {https://doi.org/10.1016/j.apsusc.2004.09.091} {\bibfield  {journal}
  {\bibinfo  {journal} {Appl. Surf. Sci.}\ }\textbf {\bibinfo {volume} {241}},\
  \bibinfo {pages} {43} (\bibinfo {year} {2005})}\BibitemShut {NoStop}%
\bibitem [{\citenamefont {Niimi}\ \emph {et~al.}(2006)\citenamefont {Niimi},
  \citenamefont {Matsui}, \citenamefont {Kambara}, \citenamefont {Tagami},
  \citenamefont {Tsukada},\ and\ \citenamefont
  {Fukuyama}}]{NiimiMatsuiKambaraEtAl2006}%
  \BibitemOpen
  \bibfield  {author} {\bibinfo {author} {\bibfnamefont {Y.}~\bibnamefont
  {Niimi}}, \bibinfo {author} {\bibfnamefont {T.}~\bibnamefont {Matsui}},
  \bibinfo {author} {\bibfnamefont {H.}~\bibnamefont {Kambara}}, \bibinfo
  {author} {\bibfnamefont {K.}~\bibnamefont {Tagami}}, \bibinfo {author}
  {\bibfnamefont {M.}~\bibnamefont {Tsukada}},\ and\ \bibinfo {author}
  {\bibfnamefont {H.}~\bibnamefont {Fukuyama}},\ }\href
  {https://doi.org/10.1103/physrevb.73.085421} {\bibfield  {journal} {\bibinfo
  {journal} {Phys. Rev. B}\ }\textbf {\bibinfo {volume} {73}},\ \bibinfo
  {pages} {085421} (\bibinfo {year} {2006})}\BibitemShut {NoStop}%
\bibitem [{\citenamefont {Kobayashi}\ \emph {et~al.}(2005)\citenamefont
  {Kobayashi}, \citenamefont {Fukui}, \citenamefont {Enoki}, \citenamefont
  {Kusakabe},\ and\ \citenamefont {Kaburagi}}]{KobayashiFukuiEnokiEtAl2005}%
  \BibitemOpen
  \bibfield  {author} {\bibinfo {author} {\bibfnamefont {Y.}~\bibnamefont
  {Kobayashi}}, \bibinfo {author} {\bibfnamefont {K.}~\bibnamefont {Fukui}},
  \bibinfo {author} {\bibfnamefont {T.}~\bibnamefont {Enoki}}, \bibinfo
  {author} {\bibfnamefont {K.}~\bibnamefont {Kusakabe}},\ and\ \bibinfo
  {author} {\bibfnamefont {Y.}~\bibnamefont {Kaburagi}},\ }\href
  {https://doi.org/10.1103/physrevb.71.193406} {\bibfield  {journal} {\bibinfo
  {journal} {Phys. Rev. B}\ }\textbf {\bibinfo {volume} {71}},\ \bibinfo
  {pages} {193406} (\bibinfo {year} {2005})}\BibitemShut {NoStop}%
\bibitem [{\citenamefont {Han}\ \emph {et~al.}(2007)\citenamefont {Han},
  \citenamefont {\"Ozyilmaz}, \citenamefont {Zhang},\ and\ \citenamefont
  {Kim}}]{Han2007}%
  \BibitemOpen
  \bibfield  {author} {\bibinfo {author} {\bibfnamefont {M.~Y.}\ \bibnamefont
  {Han}}, \bibinfo {author} {\bibfnamefont {B.}~\bibnamefont {\"Ozyilmaz}},
  \bibinfo {author} {\bibfnamefont {Y.}~\bibnamefont {Zhang}},\ and\ \bibinfo
  {author} {\bibfnamefont {P.}~\bibnamefont {Kim}},\ }\href
  {https://doi.org/10.1103/PhysRevLett.98.206805} {\bibfield  {journal}
  {\bibinfo  {journal} {Phys. Rev. Lett.}\ }\textbf {\bibinfo {volume} {98}},\
  \bibinfo {pages} {206805} (\bibinfo {year} {2007})}\BibitemShut {NoStop}%
\bibitem [{\citenamefont {Stampfer}\ \emph {et~al.}(2008)\citenamefont
  {Stampfer}, \citenamefont {Güttinger}, \citenamefont {Molitor},
  \citenamefont {Graf}, \citenamefont {Ihn},\ and\ \citenamefont
  {Ensslin}}]{Stampfer2008}%
  \BibitemOpen
  \bibfield  {author} {\bibinfo {author} {\bibfnamefont {C.}~\bibnamefont
  {Stampfer}}, \bibinfo {author} {\bibfnamefont {J.}~\bibnamefont
  {Güttinger}}, \bibinfo {author} {\bibfnamefont {F.}~\bibnamefont {Molitor}},
  \bibinfo {author} {\bibfnamefont {D.}~\bibnamefont {Graf}}, \bibinfo {author}
  {\bibfnamefont {T.}~\bibnamefont {Ihn}},\ and\ \bibinfo {author}
  {\bibfnamefont {K.}~\bibnamefont {Ensslin}},\ }\href
  {https://doi.org/10.1063/1.2827188} {\bibfield  {journal} {\bibinfo
  {journal} {Appl. Phys. Lett.}\ }\textbf {\bibinfo {volume} {92}},\ \bibinfo
  {pages} {012102} (\bibinfo {year} {2008})}\BibitemShut {NoStop}%
\bibitem [{\citenamefont {Puddy}\ \emph {et~al.}(2011)\citenamefont {Puddy},
  \citenamefont {Scard}, \citenamefont {Tyndall}, \citenamefont {Connolly},
  \citenamefont {Smith}, \citenamefont {Jones}, \citenamefont {Lombardo},
  \citenamefont {Ferrari},\ and\ \citenamefont {Buitelaar}}]{Puddy2011}%
  \BibitemOpen
  \bibfield  {author} {\bibinfo {author} {\bibfnamefont {R.~K.}\ \bibnamefont
  {Puddy}}, \bibinfo {author} {\bibfnamefont {P.~H.}\ \bibnamefont {Scard}},
  \bibinfo {author} {\bibfnamefont {D.}~\bibnamefont {Tyndall}}, \bibinfo
  {author} {\bibfnamefont {M.~R.}\ \bibnamefont {Connolly}}, \bibinfo {author}
  {\bibfnamefont {C.~G.}\ \bibnamefont {Smith}}, \bibinfo {author}
  {\bibfnamefont {G.~A.~C.}\ \bibnamefont {Jones}}, \bibinfo {author}
  {\bibfnamefont {A.}~\bibnamefont {Lombardo}}, \bibinfo {author}
  {\bibfnamefont {A.~C.}\ \bibnamefont {Ferrari}},\ and\ \bibinfo {author}
  {\bibfnamefont {M.~R.}\ \bibnamefont {Buitelaar}},\ }\href
  {https://doi.org/10.1063/1.3573802} {\bibfield  {journal} {\bibinfo
  {journal} {Appl. Phys. Lett.}\ }\textbf {\bibinfo {volume} {98}},\ \bibinfo
  {pages} {133120} (\bibinfo {year} {2011})}\BibitemShut {NoStop}%
\bibitem [{\citenamefont {Magda}\ \emph {et~al.}(2014)\citenamefont {Magda},
  \citenamefont {Jin}, \citenamefont {Hagym{\'{a}}si}, \citenamefont
  {Vancs{\'{o}}}, \citenamefont {Osv{\'{a}}th}, \citenamefont {Nemes-Incze},
  \citenamefont {Hwang}, \citenamefont {Bir{\'{o}}},\ and\ \citenamefont
  {Tapaszt{\'{o}}}}]{Magda2014}%
  \BibitemOpen
  \bibfield  {author} {\bibinfo {author} {\bibfnamefont {G.~Z.}\ \bibnamefont
  {Magda}}, \bibinfo {author} {\bibfnamefont {X.}~\bibnamefont {Jin}}, \bibinfo
  {author} {\bibfnamefont {I.}~\bibnamefont {Hagym{\'{a}}si}}, \bibinfo
  {author} {\bibfnamefont {P.}~\bibnamefont {Vancs{\'{o}}}}, \bibinfo {author}
  {\bibfnamefont {Z.}~\bibnamefont {Osv{\'{a}}th}}, \bibinfo {author}
  {\bibfnamefont {P.}~\bibnamefont {Nemes-Incze}}, \bibinfo {author}
  {\bibfnamefont {C.}~\bibnamefont {Hwang}}, \bibinfo {author} {\bibfnamefont
  {L.~P.}\ \bibnamefont {Bir{\'{o}}}},\ and\ \bibinfo {author} {\bibfnamefont
  {L.}~\bibnamefont {Tapaszt{\'{o}}}},\ }\href
  {https://doi.org/10.1038/nature13831} {\bibfield  {journal} {\bibinfo
  {journal} {Nature}\ }\textbf {\bibinfo {volume} {514}},\ \bibinfo {pages}
  {608} (\bibinfo {year} {2014})}\BibitemShut {NoStop}%
\bibitem [{\citenamefont {Talirz}\ \emph {et~al.}(2013)\citenamefont {Talirz},
  \citenamefont {Söde}, \citenamefont {Cai}, \citenamefont {Ruffieux},
  \citenamefont {Blankenburg}, \citenamefont {Jafaar}, \citenamefont {Berger},
  \citenamefont {Feng}, \citenamefont {Müllen}, \citenamefont {Passerone},
  \citenamefont {Fasel},\ and\ \citenamefont {Pignedoli}}]{Talirz2013}%
  \BibitemOpen
  \bibfield  {author} {\bibinfo {author} {\bibfnamefont {L.}~\bibnamefont
  {Talirz}}, \bibinfo {author} {\bibfnamefont {H.}~\bibnamefont {Söde}},
  \bibinfo {author} {\bibfnamefont {J.}~\bibnamefont {Cai}}, \bibinfo {author}
  {\bibfnamefont {P.}~\bibnamefont {Ruffieux}}, \bibinfo {author}
  {\bibfnamefont {S.}~\bibnamefont {Blankenburg}}, \bibinfo {author}
  {\bibfnamefont {R.}~\bibnamefont {Jafaar}}, \bibinfo {author} {\bibfnamefont
  {R.}~\bibnamefont {Berger}}, \bibinfo {author} {\bibfnamefont
  {X.}~\bibnamefont {Feng}}, \bibinfo {author} {\bibfnamefont {K.}~\bibnamefont
  {Müllen}}, \bibinfo {author} {\bibfnamefont {D.}~\bibnamefont {Passerone}},
  \bibinfo {author} {\bibfnamefont {R.}~\bibnamefont {Fasel}},\ and\ \bibinfo
  {author} {\bibfnamefont {C.~A.}\ \bibnamefont {Pignedoli}},\ }\href
  {https://doi.org/10.1021/ja311099k} {\bibfield  {journal} {\bibinfo
  {journal} {Journal of the American Chemical Society}\ }\textbf {\bibinfo
  {volume} {135}},\ \bibinfo {pages} {2060} (\bibinfo {year}
  {2013})}\BibitemShut {NoStop}%
\bibitem [{\citenamefont {Ruffieux}\ \emph {et~al.}(2016)\citenamefont
  {Ruffieux}, \citenamefont {Wang}, \citenamefont {Yang}, \citenamefont
  {Sanchez-Sanchez}, \citenamefont {Liu}, \citenamefont {Dienel}, \citenamefont
  {Talirz}, \citenamefont {Shinde}, \citenamefont {Pignedoli}, \citenamefont
  {Passerone}, \citenamefont {Fumslaff}, \citenamefont {Feng}, \citenamefont
  {Muellen},\ and\ \citenamefont {Fasel}}]{Ruffieux2016}%
  \BibitemOpen
  \bibfield  {author} {\bibinfo {author} {\bibfnamefont {P.}~\bibnamefont
  {Ruffieux}}, \bibinfo {author} {\bibfnamefont {S.}~\bibnamefont {Wang}},
  \bibinfo {author} {\bibfnamefont {B.}~\bibnamefont {Yang}}, \bibinfo {author}
  {\bibfnamefont {C.}~\bibnamefont {Sanchez-Sanchez}}, \bibinfo {author}
  {\bibfnamefont {J.}~\bibnamefont {Liu}}, \bibinfo {author} {\bibfnamefont
  {T.}~\bibnamefont {Dienel}}, \bibinfo {author} {\bibfnamefont
  {L.}~\bibnamefont {Talirz}}, \bibinfo {author} {\bibfnamefont
  {P.}~\bibnamefont {Shinde}}, \bibinfo {author} {\bibfnamefont {C.~A.}\
  \bibnamefont {Pignedoli}}, \bibinfo {author} {\bibfnamefont {D.}~\bibnamefont
  {Passerone}}, \bibinfo {author} {\bibfnamefont {T.}~\bibnamefont {Fumslaff}},
  \bibinfo {author} {\bibfnamefont {X.}~\bibnamefont {Feng}}, \bibinfo {author}
  {\bibfnamefont {K.}~\bibnamefont {Muellen}},\ and\ \bibinfo {author}
  {\bibfnamefont {R.}~\bibnamefont {Fasel}},\ }\href
  {https://doi.org/10.1038/nature17151} {\bibfield  {journal} {\bibinfo
  {journal} {Nature}\ }\textbf {\bibinfo {volume} {531}},\ \bibinfo {pages}
  {489} (\bibinfo {year} {2016})}\BibitemShut {NoStop}%
\bibitem [{\citenamefont {Yang}\ \emph {et~al.}(2010)\citenamefont {Yang},
  \citenamefont {Zhang}, \citenamefont {Wang}, \citenamefont {Shi},
  \citenamefont {Shi}, \citenamefont {Gao}, \citenamefont {Wang},\ and\
  \citenamefont {Zhang}}]{Yang2010}%
  \BibitemOpen
  \bibfield  {author} {\bibinfo {author} {\bibfnamefont {R.}~\bibnamefont
  {Yang}}, \bibinfo {author} {\bibfnamefont {L.}~\bibnamefont {Zhang}},
  \bibinfo {author} {\bibfnamefont {Y.}~\bibnamefont {Wang}}, \bibinfo {author}
  {\bibfnamefont {Z.}~\bibnamefont {Shi}}, \bibinfo {author} {\bibfnamefont
  {D.}~\bibnamefont {Shi}}, \bibinfo {author} {\bibfnamefont {H.}~\bibnamefont
  {Gao}}, \bibinfo {author} {\bibfnamefont {E.}~\bibnamefont {Wang}},\ and\
  \bibinfo {author} {\bibfnamefont {G.}~\bibnamefont {Zhang}},\ }\href
  {https://doi.org/10.1002/adma.201000618} {\bibfield  {journal} {\bibinfo
  {journal} {Adv. Mater.}\ }\textbf {\bibinfo {volume} {22}},\ \bibinfo {pages}
  {4014} (\bibinfo {year} {2010})}\BibitemShut {NoStop}%
\bibitem [{\citenamefont {Diankov}, \citenamefont {Neumann},\ and\
  \citenamefont {Goldhaber-Gordon}(2013)}]{Diankov2013}%
  \BibitemOpen
  \bibfield  {author} {\bibinfo {author} {\bibfnamefont {G.}~\bibnamefont
  {Diankov}}, \bibinfo {author} {\bibfnamefont {M.}~\bibnamefont {Neumann}},\
  and\ \bibinfo {author} {\bibfnamefont {D.}~\bibnamefont {Goldhaber-Gordon}},\
  }\href {https://doi.org/10.1021/nn304903m} {\bibfield  {journal} {\bibinfo
  {journal} {ACS Nano}\ }\textbf {\bibinfo {volume} {7}},\ \bibinfo {pages}
  {1324} (\bibinfo {year} {2013})}\BibitemShut {NoStop}%
\bibitem [{\citenamefont {Hug}\ \emph {et~al.}(2017)\citenamefont {Hug},
  \citenamefont {Zihlmann}, \citenamefont {Rehmann}, \citenamefont {Kalyoncu},
  \citenamefont {Camenzind}, \citenamefont {Marot}, \citenamefont {Watanabe},
  \citenamefont {Taniguchi},\ and\ \citenamefont {Zumb{\"{u}}hl}}]{Hug2017}%
  \BibitemOpen
  \bibfield  {author} {\bibinfo {author} {\bibfnamefont {D.}~\bibnamefont
  {Hug}}, \bibinfo {author} {\bibfnamefont {S.}~\bibnamefont {Zihlmann}},
  \bibinfo {author} {\bibfnamefont {M.~K.}\ \bibnamefont {Rehmann}}, \bibinfo
  {author} {\bibfnamefont {Y.~B.}\ \bibnamefont {Kalyoncu}}, \bibinfo {author}
  {\bibfnamefont {T.~N.}\ \bibnamefont {Camenzind}}, \bibinfo {author}
  {\bibfnamefont {L.}~\bibnamefont {Marot}}, \bibinfo {author} {\bibfnamefont
  {K.}~\bibnamefont {Watanabe}}, \bibinfo {author} {\bibfnamefont
  {T.}~\bibnamefont {Taniguchi}},\ and\ \bibinfo {author} {\bibfnamefont
  {D.~M.}\ \bibnamefont {Zumb{\"{u}}hl}},\ }\href
  {https://doi.org/10.1038/s41699-017-0021-7} {\bibfield  {journal} {\bibinfo
  {journal} {npj 2D Mater. Appl.}\ }\textbf {\bibinfo {volume} {1}},\ \bibinfo
  {pages} {21} (\bibinfo {year} {2017})}\BibitemShut {NoStop}%
\bibitem [{\citenamefont {Matsui}\ \emph {et~al.}(2019)\citenamefont {Matsui},
  \citenamefont {Sato}, \citenamefont {Kita}, \citenamefont {Amend},\ and\
  \citenamefont {Fukuyama}}]{Matsui2019}%
  \BibitemOpen
  \bibfield  {author} {\bibinfo {author} {\bibfnamefont {T.}~\bibnamefont
  {Matsui}}, \bibinfo {author} {\bibfnamefont {H.}~\bibnamefont {Sato}},
  \bibinfo {author} {\bibfnamefont {K.}~\bibnamefont {Kita}}, \bibinfo {author}
  {\bibfnamefont {A.~E.~B.}\ \bibnamefont {Amend}},\ and\ \bibinfo {author}
  {\bibfnamefont {H.}~\bibnamefont {Fukuyama}},\ }\href
  {https://doi.org/10.1021/acs.jpcc.9b06885} {\bibfield  {journal} {\bibinfo
  {journal} {J. Phys. Chem. C}\ }\textbf {\bibinfo {volume} {123}},\ \bibinfo
  {pages} {22665} (\bibinfo {year} {2019})}\BibitemShut {NoStop}%
\bibitem [{\citenamefont {Amend}\ \emph {et~al.}(2018)\citenamefont {Amend},
  \citenamefont {Matsui}, \citenamefont {Sato},\ and\ \citenamefont
  {Fukuyama}}]{Amend2018}%
  \BibitemOpen
  \bibfield  {author} {\bibinfo {author} {\bibfnamefont {A.~E.~B.}\
  \bibnamefont {Amend}}, \bibinfo {author} {\bibfnamefont {T.}~\bibnamefont
  {Matsui}}, \bibinfo {author} {\bibfnamefont {H.}~\bibnamefont {Sato}},\ and\
  \bibinfo {author} {\bibfnamefont {H.}~\bibnamefont {Fukuyama}},\ }\href
  {https://doi.org/10.1380/ejssnt.2018.72} {\bibfield  {journal} {\bibinfo
  {journal} {e-J. Surf. Sci. Nanotechnol.}\ }\textbf {\bibinfo {volume} {16}},\
  \bibinfo {pages} {72} (\bibinfo {year} {2018})}\BibitemShut {NoStop}%
\bibitem [{\citenamefont {Ochi}\ \emph {et~al.}()\citenamefont {Ochi},
  \citenamefont {Kamada}, \citenamefont {Yokosawa}, \citenamefont {Mukai},
  \citenamefont {Yoshinobu},\ and\ \citenamefont {Matsui}}]{Ochi2022}%
  \BibitemOpen
  \bibfield  {author} {\bibinfo {author} {\bibfnamefont {T.}~\bibnamefont
  {Ochi}}, \bibinfo {author} {\bibfnamefont {M.}~\bibnamefont {Kamada}},
  \bibinfo {author} {\bibfnamefont {T.}~\bibnamefont {Yokosawa}}, \bibinfo
  {author} {\bibfnamefont {K.}~\bibnamefont {Mukai}}, \bibinfo {author}
  {\bibfnamefont {J.}~\bibnamefont {Yoshinobu}},\ and\ \bibinfo {author}
  {\bibfnamefont {T.}~\bibnamefont {Matsui}},\ }\href@noop {} {\ }\bibinfo
  {note} {{in preparation}}\BibitemShut {NoStop}%
\bibitem [{\citenamefont {Matsui}\ \emph {et~al.}()\citenamefont {Matsui},
  \citenamefont {Sato}, \citenamefont {Amend},\ and\ \citenamefont
  {Fukuyama}}]{Matsui20XX}%
  \BibitemOpen
  \bibfield  {author} {\bibinfo {author} {\bibfnamefont {T.}~\bibnamefont
  {Matsui}}, \bibinfo {author} {\bibfnamefont {H.}~\bibnamefont {Sato}},
  \bibinfo {author} {\bibfnamefont {A.~E.~B.}\ \bibnamefont {Amend}},\ and\
  \bibinfo {author} {\bibfnamefont {H.}~\bibnamefont {Fukuyama}},\ }\href@noop
  {} {\ }\bibinfo {note} {{in preparation}}\BibitemShut {NoStop}%
\bibitem [{\citenamefont {Shi}\ \emph {et~al.}(2011)\citenamefont {Shi},
  \citenamefont {Yang}, \citenamefont {Zhang}, \citenamefont {Wang},
  \citenamefont {Liu}, \citenamefont {Shi}, \citenamefont {Wang},\ and\
  \citenamefont {Zhang}}]{Shi2011}%
  \BibitemOpen
  \bibfield  {author} {\bibinfo {author} {\bibfnamefont {Z.}~\bibnamefont
  {Shi}}, \bibinfo {author} {\bibfnamefont {R.}~\bibnamefont {Yang}}, \bibinfo
  {author} {\bibfnamefont {L.}~\bibnamefont {Zhang}}, \bibinfo {author}
  {\bibfnamefont {Y.}~\bibnamefont {Wang}}, \bibinfo {author} {\bibfnamefont
  {D.}~\bibnamefont {Liu}}, \bibinfo {author} {\bibfnamefont {D.}~\bibnamefont
  {Shi}}, \bibinfo {author} {\bibfnamefont {E.}~\bibnamefont {Wang}},\ and\
  \bibinfo {author} {\bibfnamefont {G.}~\bibnamefont {Zhang}},\ }\href
  {https://doi.org/10.1002/adma.201100633} {\bibfield  {journal} {\bibinfo
  {journal} {Adv. Mater.}\ }\textbf {\bibinfo {volume} {23}},\ \bibinfo {pages}
  {3061} (\bibinfo {year} {2011})}\BibitemShut {NoStop}%
\bibitem [{\citenamefont {Horcas}\ \emph {et~al.}(2007)\citenamefont {Horcas},
  \citenamefont {Fern{\'{a}}ndez}, \citenamefont
  {G{\'{o}}mez-Rodr{\'{\i}}guez}, \citenamefont {Colchero}, \citenamefont
  {G{\'{o}}mez-Herrero},\ and\ \citenamefont {Baro}}]{WSxM}%
  \BibitemOpen
  \bibfield  {author} {\bibinfo {author} {\bibfnamefont {I.}~\bibnamefont
  {Horcas}}, \bibinfo {author} {\bibfnamefont {R.}~\bibnamefont
  {Fern{\'{a}}ndez}}, \bibinfo {author} {\bibfnamefont {J.~M.}\ \bibnamefont
  {G{\'{o}}mez-Rodr{\'{\i}}guez}}, \bibinfo {author} {\bibfnamefont
  {J.}~\bibnamefont {Colchero}}, \bibinfo {author} {\bibfnamefont
  {J.}~\bibnamefont {G{\'{o}}mez-Herrero}},\ and\ \bibinfo {author}
  {\bibfnamefont {A.~M.}\ \bibnamefont {Baro}},\ }\href
  {https://doi.org/10.1063/1.2432410} {\bibfield  {journal} {\bibinfo
  {journal} {Rev. Sci. Instrum.}\ }\textbf {\bibinfo {volume} {78}},\ \bibinfo
  {pages} {013705} (\bibinfo {year} {2007})}\BibitemShut {NoStop}%
\end{thebibliography}

\providecommand{\noopsort}[1]{}\providecommand{\singleletter}[1]{#1}%

\end{document}